\documentclass[journal]{IEEEtran}
\usepackage{xcolor,soul,framed} %,caption

\colorlet{shadecolor}{yellow}
\usepackage[pdftex]{graphicx}
\graphicspath{{../pdf/}{../jpeg/}}
\DeclareGraphicsExtensions{.pdf,.jpeg,.png}

\usepackage[cmex10]{amsmath}
\usepackage{array}
\usepackage{mdwmath}
\usepackage{mdwtab}
\usepackage{eqparbox}
\usepackage{url}
\usepackage{upgreek}
\usepackage{xcolor}
\usepackage{xspace}

\usepackage[acronym,shortcuts]{glossaries}

\newacronym{MCF}{MCF}{multi-core fiber}
\newacronym{MMF}{MMF}{multi-mode fibers}
\newacronym{NCPA}{NCPA}{non-common path aberration}
\newacronym{NIR}{NIR}{near infra-red}

\newcommand{\comment}[1]{}
\newcommand{\clr}[2]{{\color{#1}#2}}
\renewcommand{\clr}[2]{{\color{black}#2}}

\begin{document}
\bstctlcite{IEEEexample:BSTcontrol}
    \title{Astronomical Applications of Multi-Core Fiber Technology}
    % I reckon title could use work if you have flexibility with it. something like: "Moulded Starlight: overview of Astronomical uses of Multi-core fiber technologies" etc.
    % An Overview of the Astronomical Applications of Multi-Core Fiber Technology
    % Multi-Core Fibres: Enabling astronomical discovery
    % Multi-Core Fibres: Enabling next-generation of astronomical instrumentation
    % Multi-Core Fibres and their astronomical applications
    % The broad range of applications of MCF technology in astronomy
  \author{Nemanja Jovanovic, Robert J. Harris and Nick Cvetojevic%

  \thanks{Manuscript received October 1st, 2019.}
  \thanks{N. Jovanovic is with The California Institute of Technology (Caltech), 1200 E. California Blvd., MC11-17, Pasadena, 91125, USA (e-mail: jovanovic.nem@gmail.com).}% <-this % stops a space
  \thanks{R. J. Harris is with the Landessternwarte~(LSW), Zentrum f\"{u}r Astronomie der Universit\"{a}t Heidelberg, K\"{o}nigstuhl 12, 69117~Heidelberg, Germany} 
  \thanks{N. Cvetojevic is with the Laboratoire Lagrange, Observatoire de la C\^{o}te d'Azur, Universit\'{e} C\^{o}te d'Azur, 06304 Nice, France} }

%\markboth{IEEE TRANSACTIONS ON MICROWAVE THEORY AND TECHNIQUES, VOL.~60, NO.~12, DECEMBER~2012
%}{Roberg \MakeLowercase{\textit{et al.}}: High-Efficiency Diode and Transistor Rectifiers}

% ====================================================================
\maketitle

% === ABSTRACT ====================================================================
% =================================================================================
\begin{abstract}
%\boldmath

Optical fibers have altered astronomical instrument design by allowing for a complex, often large instrument to be mounted in a remote and stable location with respect to the telescope. The fibers also enable the possibility to rearrange the signal from a focal plane to form a psuedo-slit at the entrance to a spectrograph, optimizing the detector usage and enabling the study of hundreds of thousands of stars or galaxies simultaneously. Multi-core fibers in particular offer several favorable properties with respect to traditional fibers: 1) the separation between single-mode cores is greatly reduced and highly regular with respect to free standing fibers, 2) they offer a monolithic package with multi-fiber capabilities and 3) they operate at the diffraction limit. These properties have enabled the realization of single component photonic lanterns, highly simplified fiber Bragg gratings, and advanced fiber mode scramblers. In addition, the precise grid of cores has enabled the design of efficient single-mode fiber integral field units for spectroscopy. In this paper, we provide an overview of the broad range of applications enabled by multi-core fiber technology in astronomy and outline future areas of development.

\end{abstract}

% === KEYWORDS ============================================================
% =========================================================================

% \fi
\IEEEpeerreviewmaketitle

% ====================================================================
% ====================================================================
% ====================================================================

%%--------------------------Section 1---------------------------------%%
% ======================================================================
\section{Introduction}\label{sec:I} 

\IEEEPARstart{S}{pectroscopy} is a critical diagnostic tool in astronomy and can be used to analyze the chemical composition, velocity, and distance to an astrophysical object. Astronomical spectrographs were originally attached directly to the telescope structure, which meant that they experienced a varying gravity vector throughout an observation as the telescope tracked the target. This placed tight constraints on the mechanical design in order to ensure the spectrograph did not flex and distort during observations. An optical fiber could be used to feed the spectrograph allowing it to be located remote to the telescope, but it wasn't until the late $1970$'s when \acf{MMF} had matured that they were seriously considered for such astronomical applications. Soon after, the first fiber-fed spectrographs were demonstrated~\cite{hill1980,hill1988}. 

\clr{red}{Another advantage of using a fiber feed is that many fibers can be routed to a spectrograph and used to form a psuedo-slit, enabling multi-object spectroscopy of up to 1000s of objects at once. Multi-object spectroscopy can also be carried out using mask slit plates, but fiber positioners conveniently allow for the input distribution of fibers at the focal plane to be reconfigured for each telescope pointing while the output is reformatted into a linear psuedo slit at the spectrograph, optimizing the use of detector pixels. The 2dF Galaxy Redshift Survey team, for example, exploited this possibility by feeding several spectrographs mounted at the primary focus of the Australian Astronomical Telescope (AAT) with hundreds of fibers. This boosted their observational efficiency by orders of magnitude enabling the first high-redshift maps of galaxies in the Universe to be produced~\cite{colless2001,croom2004}. Later, they replaced the spectrographs at the primary focus with a larger, more capable spectrograph called AAOmega, which was located in a room remote to the telescope, made possible due to the use of optical fibers. AAOmega increased their observational efficiency by another factor of 2~\cite{sharp2006}. The ability to use fiber feeds and locate spectrographs remotely ushered in an era of very large-scale extra-galactic surveys ($100,000+$ objects)}.

Astronomical fiber optic spectroscopy  (to distinguish from slit based spectroscopy) comes in three broad categories, with each favouring different kinds of spectrograph design. These categories are entirely determined by the type of astronomical detection one wishes to pursue. The three approaches along with example science cases are listed below.
\begin{itemize}
    \item {\textbf{Single-object:} A single astronomical target is captured by an optical fiber and the signal routed to the spectrograph. An example of a single-object spectrograph is one designed to operate at very high spectral resolution. Such an instrument requires many pixels to cover a broad range across the spectrum and is limited in the number of objects it can study at once.}
    \item {\textbf{Multi-object:} Multiple objects in the field-of-view are collected by separate fibers and routed to the spectrograph. Typically these fibers are aligned in a straight line that forms a psuedo-slit at the input to a spectrograph, with the spectra from each object forming a distinct track on the detector. This method is commonly used for surveys which plan to study spectra from numerous targets spread out across the sky.} 
    \item {\textbf{Integral field units (IFUs):} An extended object, such as a galaxy or nebula, is sampled by an array of closely-spaced optical fibers and mapped to a psuedo-slit at the input to the spectrograph. Using this method its possible to gain spectroscopic information across a region of the image plane.}  
\end{itemize}

With the approach established, one must next consider the fiber type to use, whether it be MMF or single-mode fiber (SMF) and whether single-fibers or a fiber bundle is most appropriate for the application. To date, MMFs have been extensively utilized for all three approaches owing to their ability to efficiently collect light that has been aberrated by the atmosphere. SMFs, operating at the diffraction limit offer the possibility for a highly miniaturized instrument but are also notoriously difficult to directly couple to from a large telescope aperture. Efficient SMF injection has however been demonstrated recently~\cite{jovanovic2017a} and there are several instruments being specifically designed for the single-object case including PARVI~\cite{Gibson2019} and iLocater~\cite{crepp2016}. The MODHIS instrument is currently being designed for the multi-object approach~\cite{mawet2019} and the RHEA instrument is the only SMF IFU demonstrated thus far\clr{red}{~\cite{rains2016,rains2018}}. \Ac{MCF}, which we define here as any fiber that supports numerous cores within a single cladding and is formed with tapering/fusing techniques (to distinguish from stacking fibers in a tube and gluing them in place), can be used to replace SMFs and enhance the performance of instruments designed for the three approaches outlined above. In this paper, we provide an overview of how \ac{MCF} technology has been exploited in the field of astronomy. Section II outlines a number of applications that were enabled and/or enhanced by exploiting \acp{MCF}. Section III offers insights into areas where further development is needed in \ac{MCF} technology to expand its reach into astronomical instrumentation and Sections IV provides some concluding statements.

%%--------------------------Section 2---------------------------------%%
% =================================================================================
\section{Multi-core Fiber Applications in Astronomy}\label{sec:II} 
This section provides a detailed overview of the broad range of applications of \ac{MCF} technology.

\subsection{Single Element Photonic Lanterns}
Photonic lanterns (PLs) allow for a multimode core to be converted to several single-mode cores~\cite{sergio2013,birks2015}. This device can facilitate an efficient transition as long as two criteria are met: 1) the number of modes in the single-mode end are equal to or greater than the number of modes in the multimode end and 2) the taper is long enough that the field sees an adiabatic transition. Owing to the multimode input, a PL is ideally suited to efficiently couple a low quality beam (as a result of the atmospheric turbulence) from the focus of a telescope while delivering a diffraction-limited beam to an instrument. Delivering a diffraction-limited beam has the inherent advantage that the instruments size and volume can be greatly reduced, which reduces cost and mechanical deflection, both of which are major motivators for astronomical instruments~\cite{jovanovic2016,Jovanovic2019}. In addition, other photonic technologies which operate at the diffraction-limit, such as fiber Bragg gratings (FBGs) for example, could also be exploited. 

\begin{figure*}
  \begin{center}
  \includegraphics[width=0.98\textwidth]{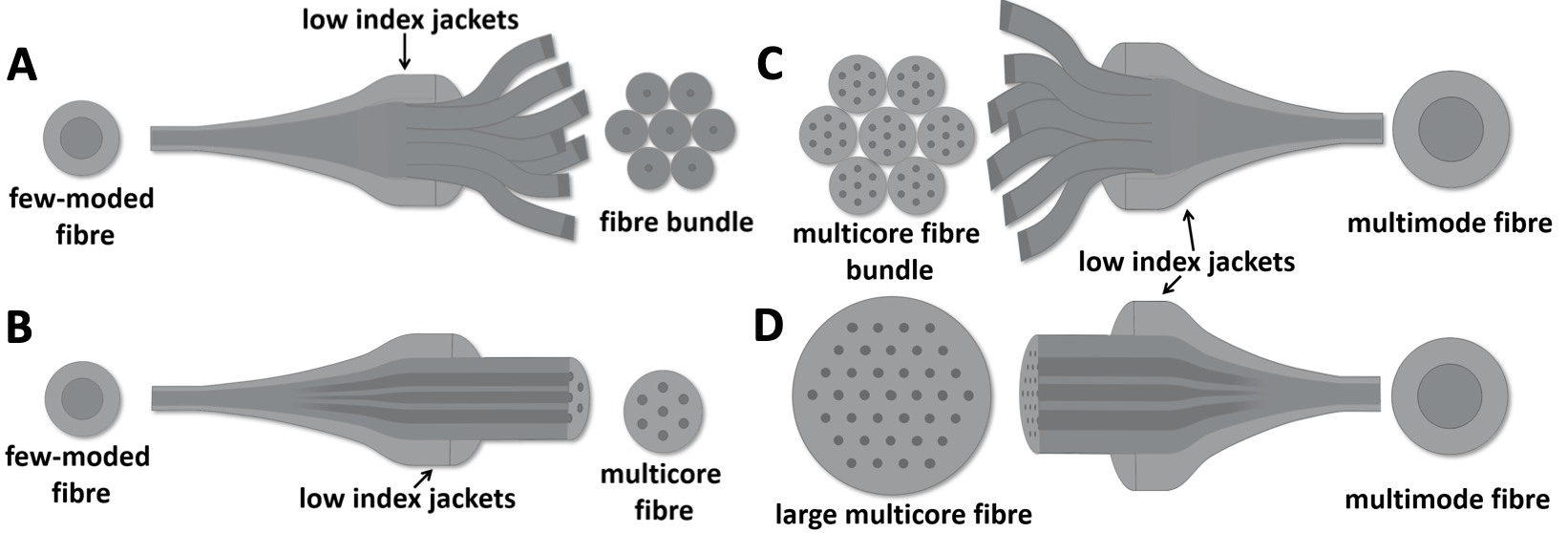}\\%3.5in
  \caption{Different varieties of Photonic Lanterns. (A) The simplest and earliest form of PLs convert the single multimode core into a number of SMFs. They are fabricated by stacking and tapering multiple SMFs and the resulting MMF section typically supports a few ($<100$) modes. (B) Alternatively, if separated SMFs are not required, a more efficient approach is to use a \ac{MCF} and taper it to a MM core. This approach scales far better than individual fibers and combining multiple MCFs (C) or using a MCF with many cores (D) allows for a truly MMF output. Reprinted/Adapted with permission from Leon-Saval et al (2017)~\cite{Leon-Saval2017} \textcopyright~The Optical Society.}\label{fig:lanterns}
  \end{center}
\end{figure*}

PLs were first realized by stacking several SMFs into a capillary and then tapering them down until the capillary formed the cladding of a new MMF and the SMFs merged so that their original cladding's formed the core of the new MMF~\cite{sergio2005} (see panel (a) in Fig.~\ref{fig:lanterns}). These early devices underwent further optimization to maximize throughput around the $1.55~\upmu$m region for a 7-port~\cite{Noordegraaf2009}, 19-port~\cite{Noordegraaf2012} and even a 61-port device~\cite{Noordegraaf2010}. An average throughput of $\sim97\%$ was achieved across several 19-port devices, which were specifically designed and optimized for an instrument called GNOSIS~\cite{trinh2013} (described in the following section).

Despite the potential of PLs, it was clear from early experiments that combining individual SMFs was not ideal due to difficulties in minimizing imperfections from manufacturing as well as handling the devices. This problem was compounded when the number of cores used was increased. An alternative approach \clr{red}{suggested by Leon-Saval et al. (2005)~\cite{sergio2005} and first experimentally demonstrated by Birks et al. (2012)~\cite{birks2012}} was to use a \ac{MCF}, place it in a low index capillary, which would eventually form the cladding for the MM section and taper that down instead (see panel (b) in Fig.~\ref{fig:lanterns}). In this demonstration, a 120 core \ac{MCF} was tapered down at both ends to realize two back-to-back PLs and hence the ability to convert multimode to single-mode and back to multimode with $<10\%$ loss. This promising method enabled PL devices to be consolidated to a single component greatly reducing the difficulty of handling the device and hence the risk of breakage during manufacturing. A \ac{MCF} based PL was tested in conjunction with a device that remapped the 2D output of the \ac{MCF} to a linear pseudo-slit (known as fan out, discussed below), on-sky behind the Canary adaptive optics system at the William Herschel Telescope~\cite{maclachlan2017}. The total throughput of the reformatting device including coupling was $>$ 50\%.

More recently, Leon-Saval et al. demonstrated an avenue to scaling PLs to very high mode counts based on \acp{MCF}. They used 7, 37-core \acp{MCF} and stacked them in a low index capillary and tapered them to create a PL~\cite{Leon-Saval2017} (see panel (c) in Fig.~\ref{fig:lanterns}). This approach enabled the equivalent of a 259 core \ac{MCF} to be used to create a high mode count PL, with a throughput of $>90\%$ (not including coupling), which will be key to applications on the extremely large telescopes.

\subsection{Compact Spectral Filters}
Fiber Bragg gratings are a mature technology that are ideally suited for spectrally filtering lines from an astronomical target. FBGs operate at the diffraction-limit which makes them difficult and inefficient to couple light to from either a seeing-limited telescope (a telescope that does not use adaptive optics (AO) to correct for the atmospheric turbulence) and/or a telescope which has a low performance AO system. The PLs described in the previous section form the ideal solution to collect light from such telescopes efficiently and exploit FBGs for spectral filtering. \clr{red}{This was recognized early on and indeed PLs and FBGs were combined to demonstrate this possibility in the founding paper~\cite{sergio2005}. The potential of this combination was later exploited by the GNOSIS instrument~\cite{trinh2013}}. 

GNOSIS was designed to collect the spectra from distant galaxies with a resolving power ($R$) of $\sim2,200$. Because of the distance to these galaxies and the expanding nature of the Universe, the spectral lines were red-shifted from the visible to the near-IR (around the $1.5~\upmu$m region). This is a region where the hydroxyl molecules in the Earth's atmosphere emit in a series of narrow bright lines. Since these lines can be much brighter than the signal from the galaxy, the light scattered by imperfections in the optics in the spectrograph from these OH lines would swamp the faint galactic signals between the lines. GNOSIS proposed to suppress the OH lines from the atmosphere before they entered the spectrograph by virtue of FBGs. Seven 19-port PLs were used to collect the light from the AATs point spread function (PSF) and convert it to 133 SMFs which fed complex FBGs that suppressed $\sim100$ OH lines in the near-IR~\cite{Bland-Hawthorn2004,Bland-Hawthorn2011}. The output SMFs were recombined with seven more PLs and fed to the IRIS2 spectrograph. Ultimately, GNOSIS successfully demonstrated that the OH lines could indeed be suppressed with a complex FBG but did not demonstrate the improvement in sensitivity expected by suppressing the OH lines, which was limited by an unoptimized spectrograph design. For this reason, a second generation spectrograph called PRAXIS was built and is currently under development~\cite{ellis2018a}. 

GNOSIS/PRAXIS used seven, 19-port lanterns which required a total of 133 individual and identical FBGs to be written. Any deviation in the properties of the FBGs from channel-to-channel would wash out the notch filters during recombination, and lead to a reduction in performance. This put stringent requirements, and significantly increased the complexity of the grating manufacturing process. 

In the process of demonstrating the first \ac{MCF} based PL described above, Birks et al. also showed that it was possible to write FBGs across all cores of a \ac{MCF} in a single step~\cite{birks2012}, potentially greatly reducing the complexity of manufacturing the devices. The preliminary device demonstrated in that work highlighted one key issue of manufacturing FBGs in \acp{MCF}, namely that when the writing laser light was focused, the cladding of the \ac{MCF} affected the beam and caused a non-uniform illumination pattern across the cores of the fiber. This resulted in FBGs with differing properties in each core, and a reduced overall performance for the spectral filter. Lindley et al. implemented one approach to correct this whereby the \ac{MCF} was placed in a capillary tube filled with index matching gel, which had one side polished flat to minimize the effect on the incident beam~\cite{Lindley2014}. This resulted in improved uniformity across all cores of a 7-core \ac{MCF}\clr{red}{, but the transmission profiles for each core were still not identical}. The same group then outlined a new configuration for their writing laser that could extend this technique to 127 cores~\cite{blandhawthorn2016}. \clr{red}{Further studies revealed that some of the residual non-uniformity in the FBGs was a result of non-uniformity in the propagation constant of the cores of the \ac{MCF}, most likely a consequence of imperfections in the preform or the drawing process~\cite{Mosley2014,ellis2018b}. These studies showed a gradient in propagation constant from the central core in the radial direction, and since the propagation constant is a function of the effective index of the mode and the FBGs resonant wavelength depends on this, the FBGs would also be shifted in wavelength in a similar manner. Until the uniformity in \ac{MCF} technology is improved, it will set the fundamental limit to what can be achieved in terms of FBG uniformity in a \ac{MCF}.} However, methods to deferentially post-tune the FBGs in individual cores of a \ac{MCF} were also demonstrated as a possibility to correct for manufacturing errors~\cite{lindley2016}. Despite these efforts, much work still remains to optimize the grating uniformity in \acp{MCF} to a level sufficient for efficient use in astronomical applications.

\subsection{Optimizing Spectrograph Feeds}
\begin{figure}
  \begin{center}
  \includegraphics[width=0.49\textwidth]{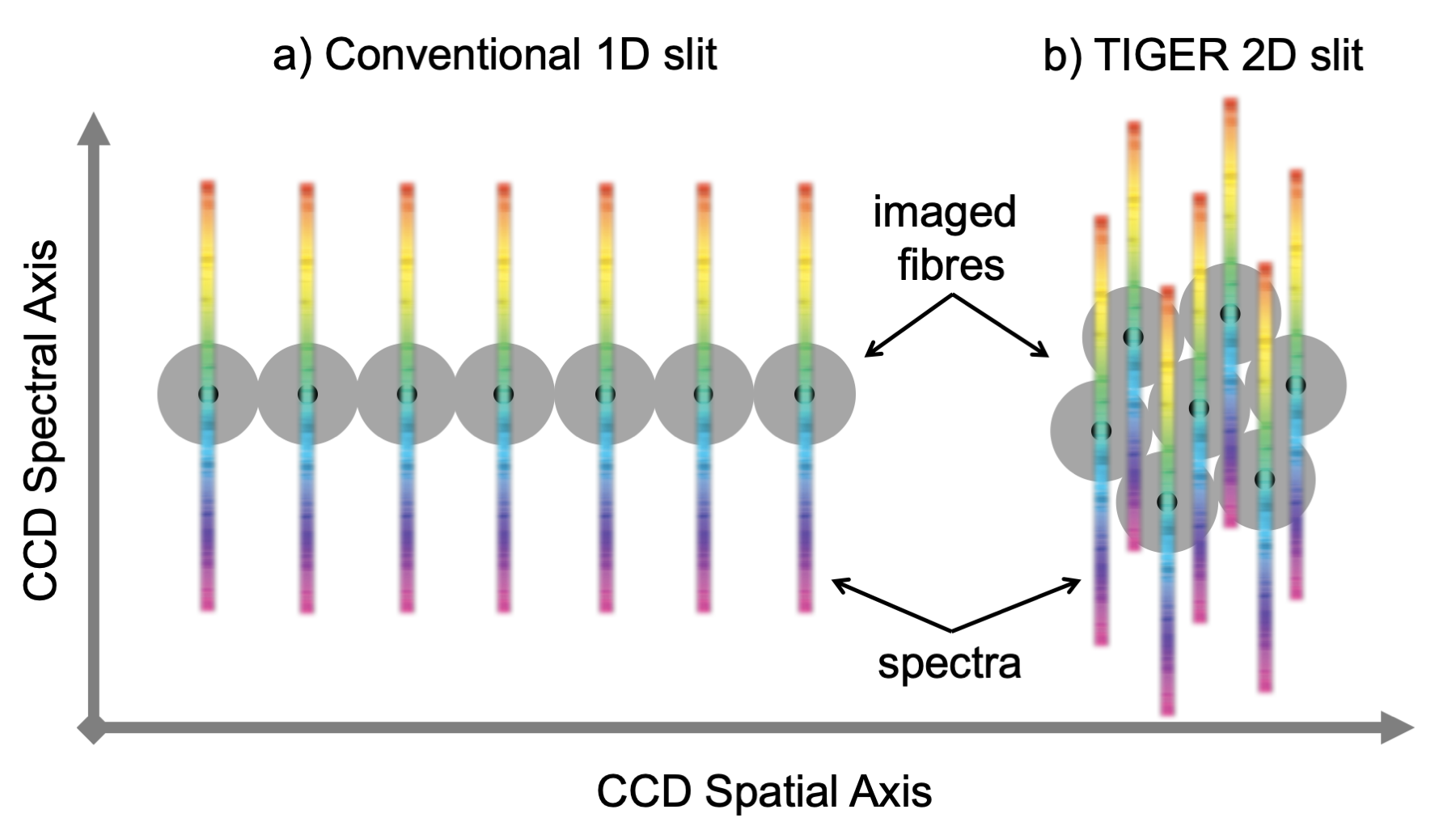}\\
  \caption{An illustration of various psuedo-slit geometries and how the dispersed spectra lie on the detector. (Left) a linear array of SMFs butted cladding-to-cladding. (Right) a hexagonal array clocked slightly so the cores do not disperse onto one another known as the Tiger concept. Credit: Image reproduced from Betters et al (2014)~\cite{betters2016}.}\label{fig:slit}
  \end{center}
\end{figure}

\acp{MCF} can offer several advantages when it comes to feeding a spectrograph as well. For example, the cladding of a SMF is typically $125~\upmu$m while the core is closer to $3-10~\upmu$m in diameter depending on the wavelength the fiber was intended to operate at. It is ideal to sample the re-imaged SMF cores on the detector inside the spectrograph with at least two pixels in the direction orthogonal to the dispersion direction. If one were to form a linear psuedo-slit from SMFs (see left panel of Fig.~\ref{fig:slit}), this implies that the distance between two adjacent SMFs will be between 83 and 25 pixels on the detector respectively. With such a large spacing between adjacent cores, this would result in a large number of un-illuminated pixels on the very costly detector. A \ac{MCF} on the other hand reduces the separations between cores from $125~\upmu$m down to as low as $15~\upmu$m in some extreme cases, vastly improving the fiber packing factor and hence the use of detector real estate.

In addition, the \ac{MCF} geometry is such that the cores are not aligned in a linear pattern but rather a square or most commonly a hexagonal grid. When using such a grid care must be taken to clock the output so no two cores are spatially aligned in the direction of dispersion (see right panel of Fig.~\ref{fig:slit}). This approach is called the ``photonic Tiger'' approach named after a spectrograph that used micro-lenses to do something very similar and has been demonstrated in several instruments to date~\cite{Leon-Saval2012,betters2014,betters2016}. In this fashion its possible to reduce the spacing between two spectral traces by even more, greatly improving the packing on the final detector. \clr{red}{It is worth mentioning that the offset in the spectral direction due by the Tiger geometry will slightly reduce the maximum length of a given free spectral range to avoid the edges falling off the detector and should be accounted for while designing the spectrograph to optimize performance.}  

Several groups have also investigated using 3D photonic waveguides fabricated by ultrafast laser inscription (ULI) to reformat the 2D fiber array at the output of a \ac{MCF} to a linear psuedo-slit where the cores in the slit are spaced close enough to cause cross coupling~\cite{spaleniak2016,theo2018}. Both groups witnessed modal noise effects from the interference within the slit, which if not controlled carefully can prohibit science cases such as precision radial velocity detection of exoplanets orbiting their host stars. Reducing the modal noise of such a device is possible through careful design and this approach can offer an alternative to the Tiger concept which may not be suitable for all applications.

\subsection{Efficient Fiber Scrambling}
Extrasolar planets, or planets around other stars, are commonly detected using the Doppler or radial velocity technique. The presence of a planet can be inferred by studying the subtle shifts in the spectral lines of the host star due to the effect of the planet pulling on its host as it orbits. Several Jupiter mass planets can induce velocity shifts in the host star of the order of a few km/s, which would correspond to a wavelength shift of the order of the resolution element of an $R\sim100,000$ spectrograph. However, lower mass planets, like Earths and super-Earths induce significantly smaller shifts ($<3$~m/s), which requires the ability to detect the motion of the spectral lines to $<1/1000^{th}$ of a resolution element. This extremely challenging task clearly requires a highly stable and precise instrument along with advanced calibration procedures.

Single-object spectrographs, which exploit MMF-feeds are typically used for this science case. This approach allows for efficient fiber coupling and enables the instrument to be located in a highly temperature stabilized room, required to reach the extreme precision's needed. However, two key limitations have been observed when utilizing MMF: namely incomplete scrambling and modal noise. Incomplete scrambling refers to persistent structure from the telescope beam seen at the output of the fiber and hence in the spectrum produced on the detector. This can be a deleterious effect because any motion in the centroid of the input beam to the fiber could be preserved at the output of the fiber and seen in the final spectrum and would appear as a motion in the barycentre of the measured line. This would obscure small frequency shifts induced by the object itself, potentially hiding the presence of a planet. To address this, scrambling techniques have been developed which include using non-circular MMFs~\cite{chazelas2010,roy2014} as well as double-scrambling with ball lenses~\cite{hunter1992}. 

In addition, due to imperfections in the fiber, light couples between modes and excites new modes as it propagates along the fiber leading to an unstable output beam that varies with wavelength and time. This effect can also be confused as a barycenter shift in the measured spectral line and is known as modal noise. To combat this, fibers are generally agitated~\cite{plavchan2013}. Although these scrambling techniques work well, they are not perfect. These imperfections are exacerbated as one moves towards the~\ac{NIR} where there are less modes in the optical fiber, amplifying the effect of modal noise~\cite{harris2016}.

\begin{figure*}
  \begin{center}
  \includegraphics[width=0.95\textwidth]{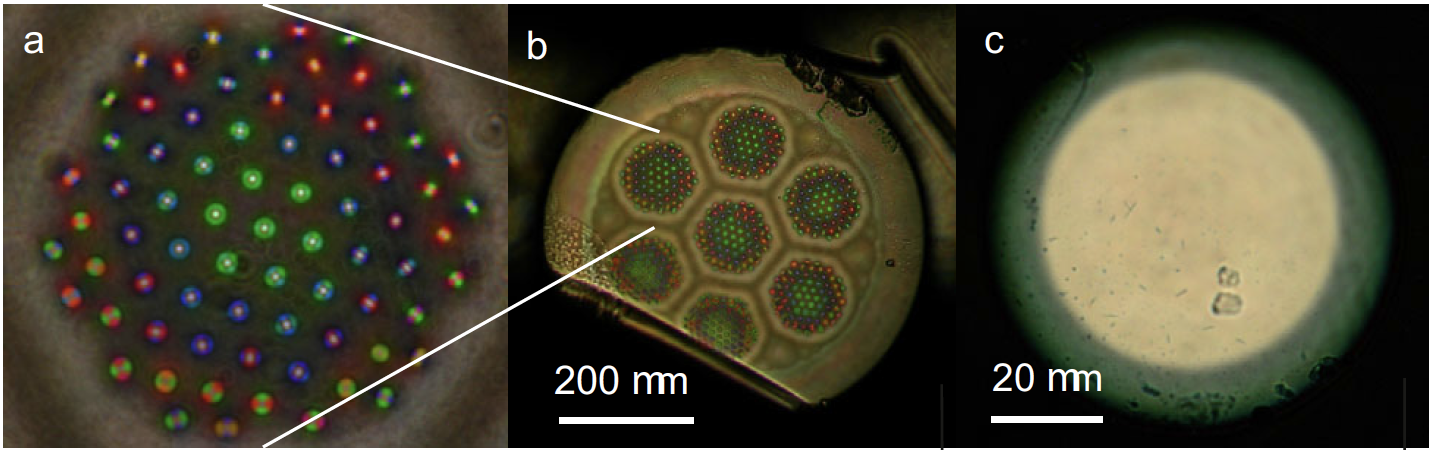}\\%3.5in
  \caption{(a) Image of the output of a 73-core \ac{MCF} with modified core diameters. (b) Image of seven 73-core \acp{MCF} fused in a single capillary tube. (c) The seven 73-core \acp{MCF} tapered to form a MM waveguide. Credit: Image reproduced from Gris-Sanchez et al (2018)~\cite{gris2018} \textcopyright.}\label{fig:scrambler}
  \end{center}
\end{figure*}

\acp{MCF} in combination with PLs offer a unique solution to these issues. Indeed, it was realized early-on that a back-to-back PL built from a \ac{MCF} could offer mode scrambling properties~\cite{birks2012}. However, because the \ac{MCF} cores were almost identical in diameter and length in that early work, the scrambling effect was minimal. To increase the scrambling, the difference between the \ac{MCF} cores was maximised, while ensuring the PL still functioned. Specifically, the \ac{MCF} core diameters were modified in such a way as to allow for higher order modes to propagate in some of the nominally single-moded cores. This can be seen in the microscope image of the end facet of one of the 73-core lanterns shown in panel (a) of Fig.~ \ref{fig:scrambler}~\cite{gris2018}. As the light propagates in different spatial modes in each of the cores it picks up a phase shift due to the different propagation constants for each mode. In addition, light moves between modes within a single core due to perturbations and imperfections so the amplitude of each core changes from input to the out. When the light from the MCF cores is recombined in the output PL, the phase and amplitude changes induced in the device act as an efficient way to scramble both phase and amplitude when it arrives at the output MMF (spectrograph psuedo-slit). These new devices proved to provide extremely efficient passive modal-scrambling performance, similar to the more conventionally used octagonal fibers, potentially offering a highly simplified solution for future extreme precision instruments that want to utilize MMFs~\cite{haynes2014,haynes2018,gris2018}.

\subsection{Enhancing Diffraction-Limited Single-Object Spectroscopy}

Operating at the diffraction-limit not only allows for the most compact instrument design possible, but it offers a highly stable PSF to the instrument because of the single mode nature of a delivery fiber, which can be calibrated to extraordinary precision~\cite{jovanovic2016,Jovanovic2019}. Being able to miniaturize the spectrograph and decouple its internal beam shape stability from that of the telescope is advantageous from the point of view of cost, stability and precision. These advantages were recognized by several groups and the first wave of diffraction-limited astronomical spectrographs are currently under construction including the PARVI instrument designed for the Palomar Observatory~\cite{Gibson2019}, iLocater~\cite{crepp2016} for the Large Binocular Telescope and the HISPEC/MODHIS concepts for the Keck and TMT telescopes~\cite{mawet2019}.  

However, directly injecting light from a large aperture ground-based telescope into a SMF is challenging. Recently, Jovanovic et al. demonstrated that with a high performance AO system, which are becoming common place at all major observatories, as well as beam shaping optics it was possible to achieve $>50\%$ coupling efficiencies in median seeing conditions~\cite{jovanovic2017a}. But minimizing \acp{NCPA} between the wavefront sensor arm and the focal plane where the fiber is located, as well as telescope structure vibrations~\cite{Kulcsar2012}, is a challenge that needs to be solved to maximize coupling into the SMFs. Accelerometers can be used to track telescope vibrations and potentially drive a feed forward loop to to take some of the work off the AO system~\cite{Gluck2017}. \acp{MCF} can be used to address the \ac{NCPA} problem.

\begin{figure}
  \begin{center}
  \includegraphics[width=0.49\textwidth]{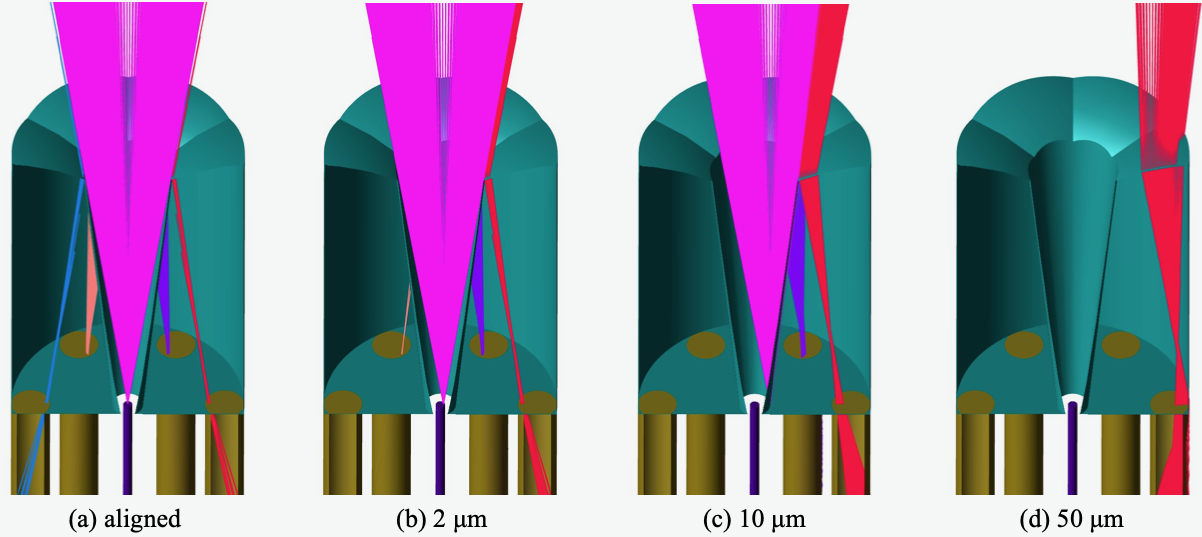}\\%3.5in
  \caption{The image shows the concept of a 3D printed micro-lens array placed on the input facet of a 7-core \ac{MCF} and how the rays couple to various cores depnding on alignment. (a) With the beam optimally aligned with the central core, (b) with the beam offset by $2~\upmu$m and light starting to couple into neighbouring cores, (c) with the beam offset by $10~\upmu$m and more light coupling into the neighbouring cores and (d) with the beam offset by $50~\upmu$m and all of the light coupling into a neighbouring core. Credit: Image reproduced from Hottinger et al (2018)~\cite{hottinger2018}.}\label{fig:TTsensor}
  \end{center}
\end{figure}

The concept of using a \ac{MCF} for \ac{NCPA} compensation, can be done in several ways. The simplest way is to place a micro-lens array (MLA) in front of a \ac{MCF}, so that the light from the science target is coupled into the central micro-lens and core. Initially there would be little to no light in the peripheral cores. As tip-tilt errors increase, they would steer light into the neighboring micro-lenses and associated cores, which would create a signal that can be used to drive a tip-tilt or deformable mirror to make a correction upstream. This was first developed using a seven core \ac{MCF}, and exploited 3D printing to place a MLA on top of all cores~\cite{Dietrich2017}. The authors were successful in demonstrating that the device generated a strong signal when tip-tilt errors were present, and could be used for wavefront sensing. More recently, the linear range and throughput of such a sensor was improved by substituting the peripheral SMFs with MMFs and removing the central micro-lens as shown in Fig.~\ref{fig:TTsensor}~\cite{hottinger2018,hottinger2019}. By removing the central micro-lens, it is possible to use the device in the focal plane of an existing system. This device was actually a fiber bundle and not a \ac{MCF} constructed from stacking and gluing individual fibers in place, but the concept could be extended to use a hybrid \ac{MCF}, which contains both single and MMF cores in future. Both of these approaches are very similar to using a pinhole for guiding, in that a hole is used to transmit the light from the science object to the spectrograph, whilst the surrounding image is used for guiding. Such a sensor would enable efficient tip/tilt and possibly low order sensing and control to help drive the control loop to maximize flux for the science fiber.

Another option to conduct single-object spectroscopy is to use a few port PL to feed the spectrograph and do the focal plane wavefront sensing instead of the \ac{MCF}. In this architecture, the few port PL improves the coupling efficiency from a low quality telescope beam with respect to a SMF while delivering diffraction-limited performance to the spectrograph. In addition, Corrigan et al. showed that it was theoretically possible to use the output signals from the fibers of the few port PL to determine the amplitude and direction of tip/tilt jitter in the beam at the input to the device~\cite{Corrigan:2016}. Norris et al. have advanced this concept by demonstrating in the laboratory that they can successfully determine the amplitude and phase of eight Zernike mode aberrations applied to a beam at the input to a 19-port PL~\cite{norris2019a}. This approach used a neural network and a training data set to build an understanding of the PL response and could be applied to any basis set of modes. Closed loop correction has not yet been demonstrated but is planned for the near future. \clr{red}{It may also be possible to substitute the bulk optic deformable mirror with a photonic mode-routing device known as a ``universal beam coupler" so the correction can be done entirely in photonics~\cite{Miller2013,birks2015}}. As an alternative to actively correcting the input electric field, a discrete waveguide array could be used downstream of the \ac{MCF} to interfere all the cores together to reconstruct the full complex field~\cite{Diab:19}. However, it is not clear how this approach can be used while trying to channel the majority of the light in at least one channel to the spectrograph. Regardless, being able to do basic wavefront sensing and control with a PL is extremely promising because it would mean that not only could the \ac{NCPA}s be compensated right at the fiber focal plane improving coupling, but it could in principle be used to funnel light to a particular core maximizing the flux.

Both methods clearly demonstrate that \ac{MCF}s have a role to play in enhancing single-object spectorscopy in the future.

\subsection{Enabling Spatially Resolved Spectroscopy}
The highly precise grid of fibers offered by a ~\ac{MCF} is ideal for building an IFU, which can be used to image across a region of the focal plane. These can be realized in two ways: 1) placing the \ac{MCF} directly in the focal plane of the telescope or 2) placing the \ac{MCF} in the focal plane behind a MLA. The latter approach offers the benefit of an improved fill factor at the expense of extra losses associated with both the throughput of the lenslets as well as coupling to the \ac{MCF}. The former approach, although simpler imposes a sparse sampling on the input focal plane field. \clr{red}{For example, a typical device may offer a mode field diameter (MFD) of $10~\upmu$m (NA$\sim0.1$--$0.14$) around $1.55~\upmu$m with a $37~\upmu$m core-to-core spacing. To achieve the highest coupling efficiency and also high spatial sampling with the IFU, the focal ratio should be adjusted such that MFD$=1.4~\lambda F\#$, where $F\#$ is the focal ratio. When this criteria is met the coupling between the diffraction-limited Airy beam of the telescope and the quasi-Gaussian mode of a SMF is optimized~\cite{echeverri2019}. This occurs for focal ratios of $4.5-5.5$ for the MFD and NA range chosen in this example, and means the diffraction-limit of the telescope or $\lambda/D\sim7.8~\upmu$m. In this case the space between the cores will be $\sim4.5~\lambda/D$}. To image the parts of the focal plane in between the cores in this case, a dither pattern would need to be used which would be an inefficient use of telescope time given how sparse this particular \ac{MCF} is. 

However, a sparsely sampled focal plane is suitable if one intends to monitor several locations in the focal plane simultaneously, rather than image it in its entirety. The REACH project uses a 7-core \ac{MCF} to conduct high-resolution spectroscopy on directly imaged exoplanets~\cite{jovanovic2017b}. The aim is to use the AO system to correct for the wavefront~\cite{jovanovic2015}, pass the light through a coronagraph to extinguish as much of the starlight as possible, align the central fiber of the \ac{MCF} with the position of the known planet in the focal plane, and route the light to the high-resolution spectrograph known as IRD~\cite{kotani2018}. The peripheral fibers of the \ac{MCF} are used to monitor the residual starlight, which is spread across the focal plane due to the atmosphere, and are also fed to the spectrograph. In this way, a spectrum is generated from the central fiber which contains the planet, residual starlight and tellurics from the atmosphere and one that contains only the residual starlight and tellurics from the peripheral fibers. By collecting these spectra contemporaneously, they can be used to remove the effect of the stellar spectrum and tellurics and improve the signal-to-noise ratio for detecting the planets spectrum. \clr{red}{An interesting advantage of using a SMF core for this application is that it suppresses unwanted starlight by a factor of 2-3 due to the difficulties of coupling the complex speckle halo into the fundamental mode, while the planet coupling is unaffected,  which is highly desirable for this science case~\cite{mawet2017}. In fact, this suppression technique is already used in interferometry, albeit in a slightly different context, where the spatial filtering the single-mode cores provide removes the sub-aperture fluctuations that contribute phase and contrast errors~\cite{coude1993}.}   

\begin{figure*}
  \begin{center}
  \includegraphics[width=0.98\textwidth]{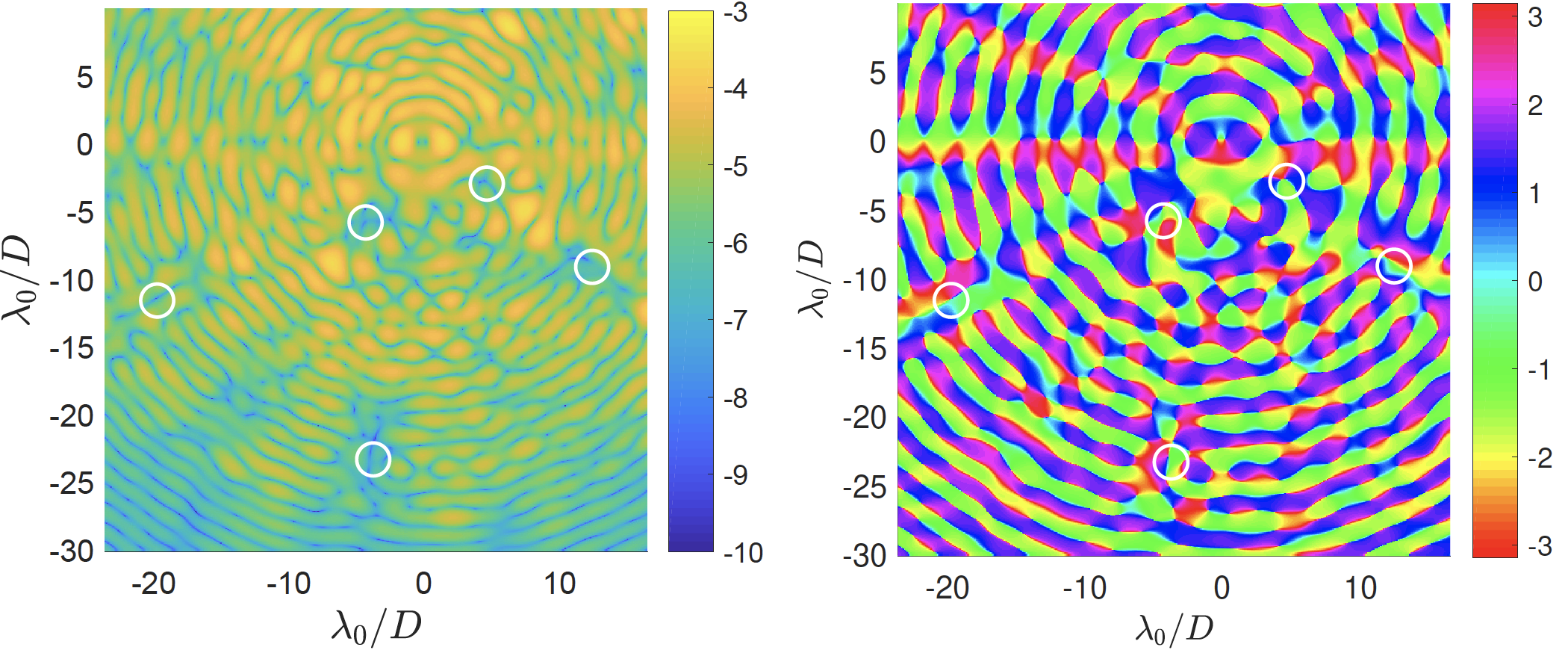}\\%3.5in
  \caption{(Left) an intensity map on a base-10 log scale and (Right) the phase of the focal plane after light has propagated through the coronagraph with wavefront control applied. The white circles indicate the position of the five micro-lenses selected for wavefront control. The intensity is clearly not zero but the phase shows a strong phase inversion at the location of the micro-lenses which strongly suppresses star light coupling into the fibers. Credit: Image reproduced from Coker et al (2019)~\cite{coker2019}.}\label{fig:FPWC}
  \end{center}
\end{figure*}

Several groups have also suggested exploiting MLAs in combination with \ac{MCF}s to build densely packed IFUs for exoplanet detection and characterization~\cite{por2018,haffert2018,coker2019}. The first approach was dubbed the ``SCAR" coronagraph and combines a pupil plane phase plate with a MLA and \ac{MCF}, to suppress starlight coupling into all fibers across the focal plane while promoting efficient coupling for a potential planet~\cite{por2018,haffert2018}. The second approach utilizes a deformable mirror to modulate the electric field and eliminate starlight leakage into select fibers~\cite{coker2019} (see Fig.~\ref{fig:FPWC}). This work has shown that wavefront control with SMFs in this manner has the potential to expand the usable spectral bandwidth over which the high contrast is maintained by a factor of 2-3. 

The RHEA instrument, currently installed at the Subaru Telescope is the first spectrograph to use a SMF-IFU to the best of the authors knowledge~\cite{rains2016,rains2018}. The IFU consists of 9 SMFs in a square geometry attached to individual micro-lenses. The plan is to replace the IFU with a \ac{MCF} in future and a 3D printed MLA on the front of fiber as demonstrated by Hottinger et al.~\cite{hottinger2018}, improving overall system efficiency. This instrument will be used to study and map the surfaces of giant resolved stars and disks.

%%--------------------------Section 3---------------------------------%%

\section{Future Applications and Developments}\label{sec:III} 
In addition to the applications \ac{MCF}s have been utilized for outlined in Section~\ref{sec:II}, \ac{MCF}s are interesting for a number of other possible applications. 

Given that the all cores are embedded in a single cladding, in close proximity to one another, the cores experience a very similar thermal environment. In addition, the cores are also path length matched. This means that \ac{MCF}s may be the ideal fiber technology for applications that require the beams in the cores to be interfered in a downstream instrument. A \ac{MCF} could be used to collect the light from a segmented pupil (with the aid of a MLA) and route it to a downstream beam combiner in a fashion akin to the pupil remapping interferometers Dragonfly/GLINT~\cite{jovanovic2012,norris2019b} and FIRST~\cite{huby2013}. These instruments are used to image objects at very small separations from the host star such as dust shells, disks and binaries. 

\ac{MCF}s could also be used as coherent fiber bundles. Richards et al. proposed two key applications of coherent fiber bundles that apply to \acp{MCF}: namely the routing of light from a focal plane to a location where it can be used for telescope guiding and/or wavefront control~\cite{richards2013}. To address both applications with a~\ac{MCF}, a MLA would need to be used to efficiently inject the light into the cores. For the guiding application, many lenslets/fibers would be needed if a high resolution image were required for guiding. For wavefront sensing, the MLA and \ac{MCF} could be placed in a pupil or focal plane. It could feed the light to a remote Shack-Hartmann or curvature wavefront sensor. Both applications are designed to decrease the foot print of a guider/wavefront sensor in the telescope focal plane allowing for the possibility of using several such systems in a crowded multi-object style focal plane. 

\ac{MCF}s based on endlessly SMFs could be interesting as well. Endlessly SMFs typically have large mode areas. \clr{red}{To couple efficiently into these, the $F\#$ would need to be made larger increasing the focal spot size. Given the larger size of the beam and fiber mode, the coupling efficiency will be less sensitive to misalignment between the fibers and the MLA}. In addition, the broad band nature of the fibers could simplify highly broadband instruments that currently require several fibers to cover the entire operating range.

Finally, one critical area that requires further development is interfacing to \ac{MCF}s. Devices dubbed ``fanouts" have been fabricated using ULI in blocks of glass~\cite{thomson2012} and optical fibers to route the light from the 2D output geometry of cores to a psuedo-slit for example. \clr{red}{The ULI approach is highly flexible and can produce devices with losses as low as $25\%$ ($\sim1$~dB) around $1.55~\upmu$m~\cite{MacLachlan2016}}. Chiral Photonics\footnote{\url{https://www.chiralphotonics.com/}} is offering fiber-based solutions for a range of \ac{MCF}s that they carry as well as for customer provided ones with insertion losses on the order of $20\%$ as well. Although promising, these losses are still too high for astronomical applications and further development is needed in this area to really make \ac{MCF}s efficient tools for astronomical instrumentation.

%%--------------------------Section 4---------------------------------%%

\section{Summary}\label{sec:IV} 
\ac{MCF} technologies have been used in many areas of astronomy to date. Most notably they have been used to design efficient PLs, enable single components FBGs, develop advanced and simplified fiber scramblers, improve wavefront control for efficient injection into single-object spectrographs and for imaging spectroscopy at the diffraction-limit. Interfacing with them to reformat the 2D output array of fibers still remains a challenge as fibers are continuously evolving but progress is being made in this area. The unique characteristics of \ac{MCF} technology may see them applied to many new areas of astronomy in the future.

\section*{Acknowledgment}

N. Cvetojevic acknowledges funding from the European Research Council (ERC) under the European Union’s Horizon 2020 research and innovation program (grant agreement CoG~-~683029). 

R. J. Harris was supported by the DFG through project 326946494, 'Novel Astronomical Instrumentation through photonic Reformatting' and, and European Commission (Fp7 Infrastructures 2012-1, OPTICON Grant 312430, WP6)

\ifCLASSOPTIONcaptionsoff
  \newpage
\fi


%\begin{thebibliography}{1}


\bibliographystyle{IEEEtran}
\bibliography{IEEEabrv,Bibliography}
%\end{thebibliography}

% ==== SWITCH OFF the BIO for submission
% ==== SWITCH OFF the BIO for submission
\begin{IEEEbiography}[{\includegraphics[width=1in,height=1.25in,clip,keepaspectratio]{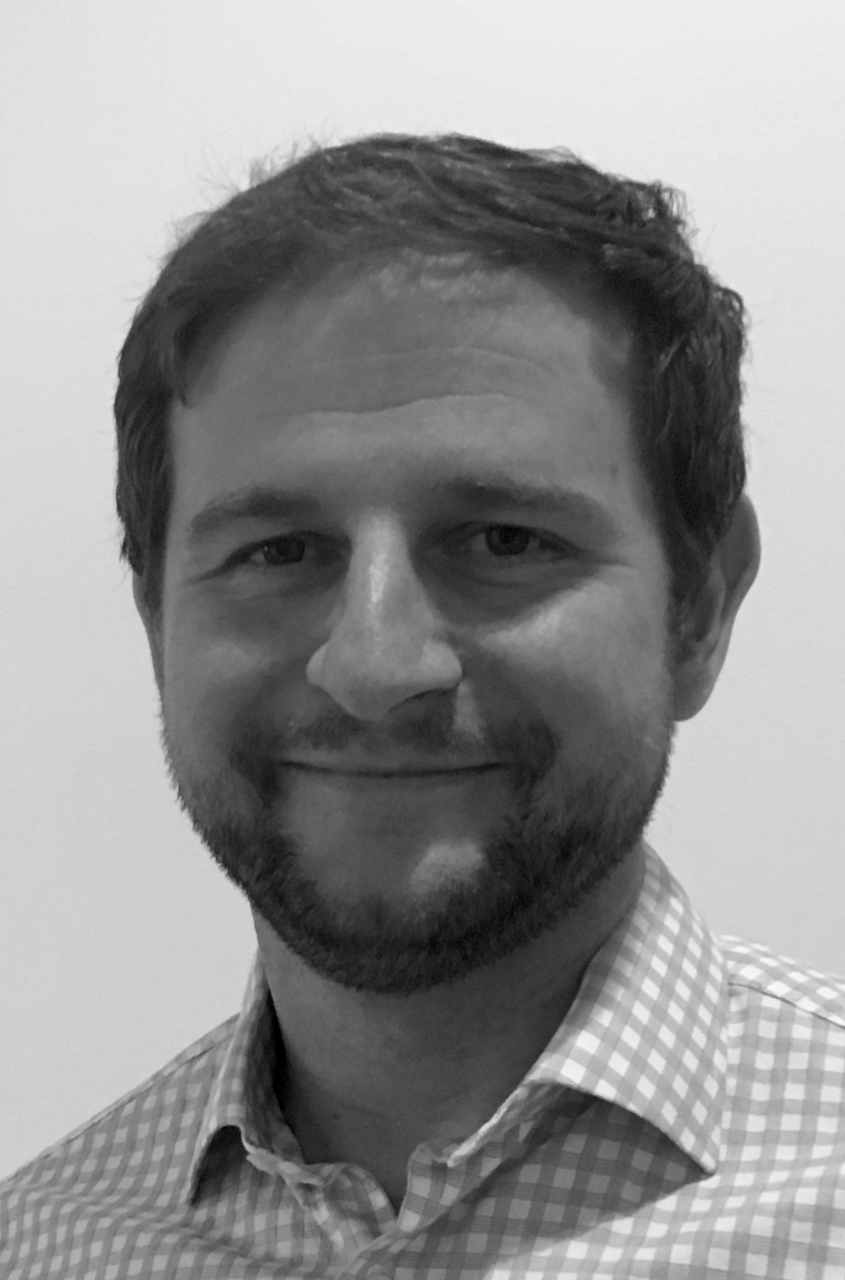}}]{Nemanja Jovanovic} received a B. Tech. Optoelectronics degree from Macquarie University, Sydney, NSW, in 2004, a honors degree and Ph.D. from the same institute in 2005 and 2010 respectively. From 2010 to 2012, he worked as a postdoc. on astrophotonics jointly appointed between Macquarie University and the Australian Astronomical Observatory, where he was involved with the development of integrated photonic components such as spectrographs, Bragg gratings, photonic lanterns and pupil remappers. He worked at Subaru Telescope between 2012 and 2017 as part of the SCExAO instrument where he demonstrated efficient injection into single mode fibers on-sky for the first time. He currently works at Caltech as a Senior Instrument Scientist developing advanced instruments for the largest ground based telescopes. His current research interests continue to be the development and implementation of photonic components to astronomical instruments.
\end{IEEEbiography}

\begin{IEEEbiography}[{\includegraphics[width=1in,height=1.25in,clip,keepaspectratio]{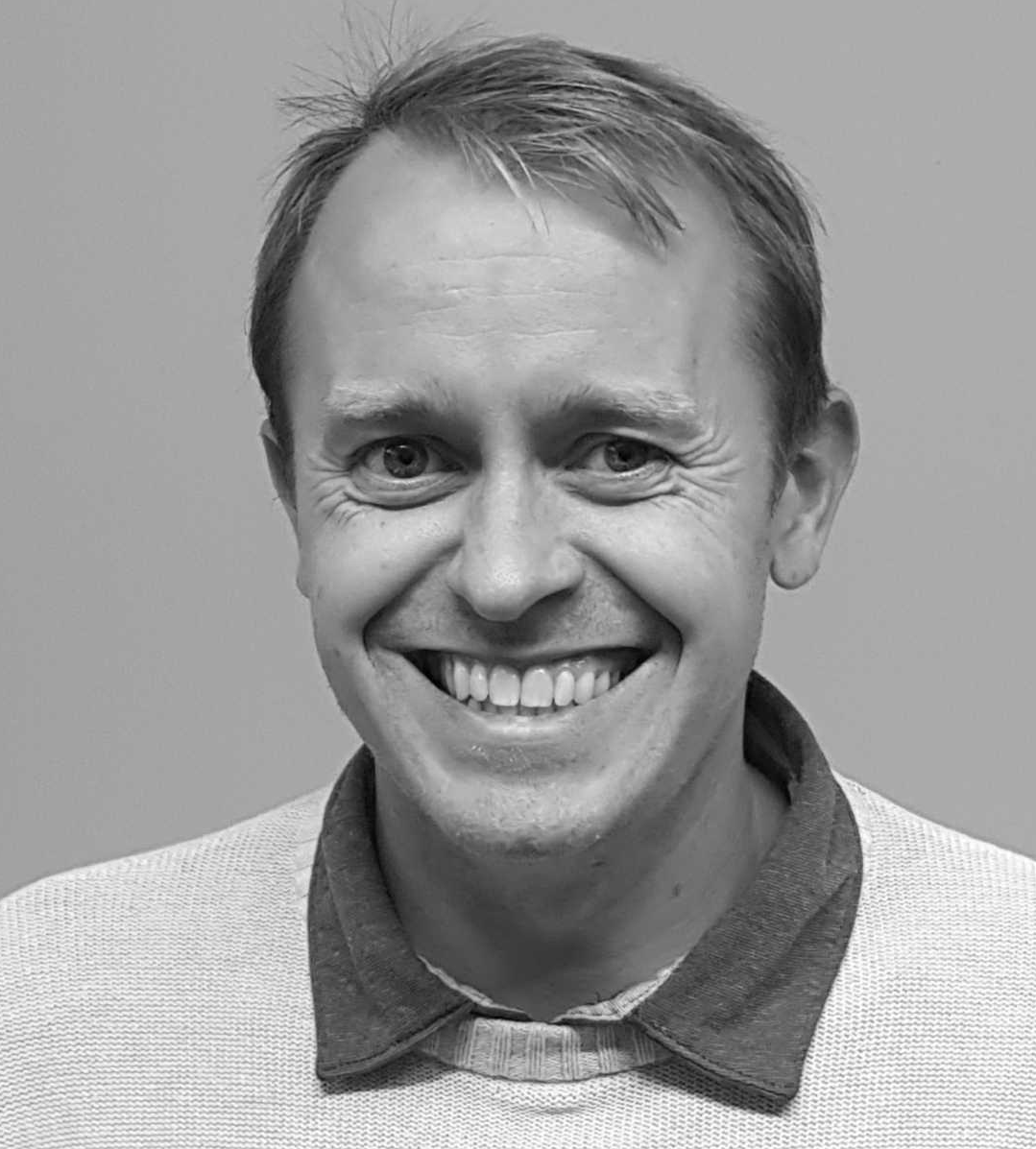}}]{Robert J. Harris}
Robert completed his PhD at Durham University in 2014, modelling astrophotonic devices and spectrographs and developing photonic reformatters for high resolution spectroscopy. Following the PhD and a STEP postdoc (again at Durham), he took up a Carl-Zeiss fellowship at the Landessternwarte, Heidelberg. Here and in his subsequent Gliese fellowship he has been developing astrophotonic components and instruments along with a small but growing group.
\end{IEEEbiography}%replace with your info please

\begin{IEEEbiography}[{\includegraphics[width=1in,height=1.25in,clip,keepaspectratio]{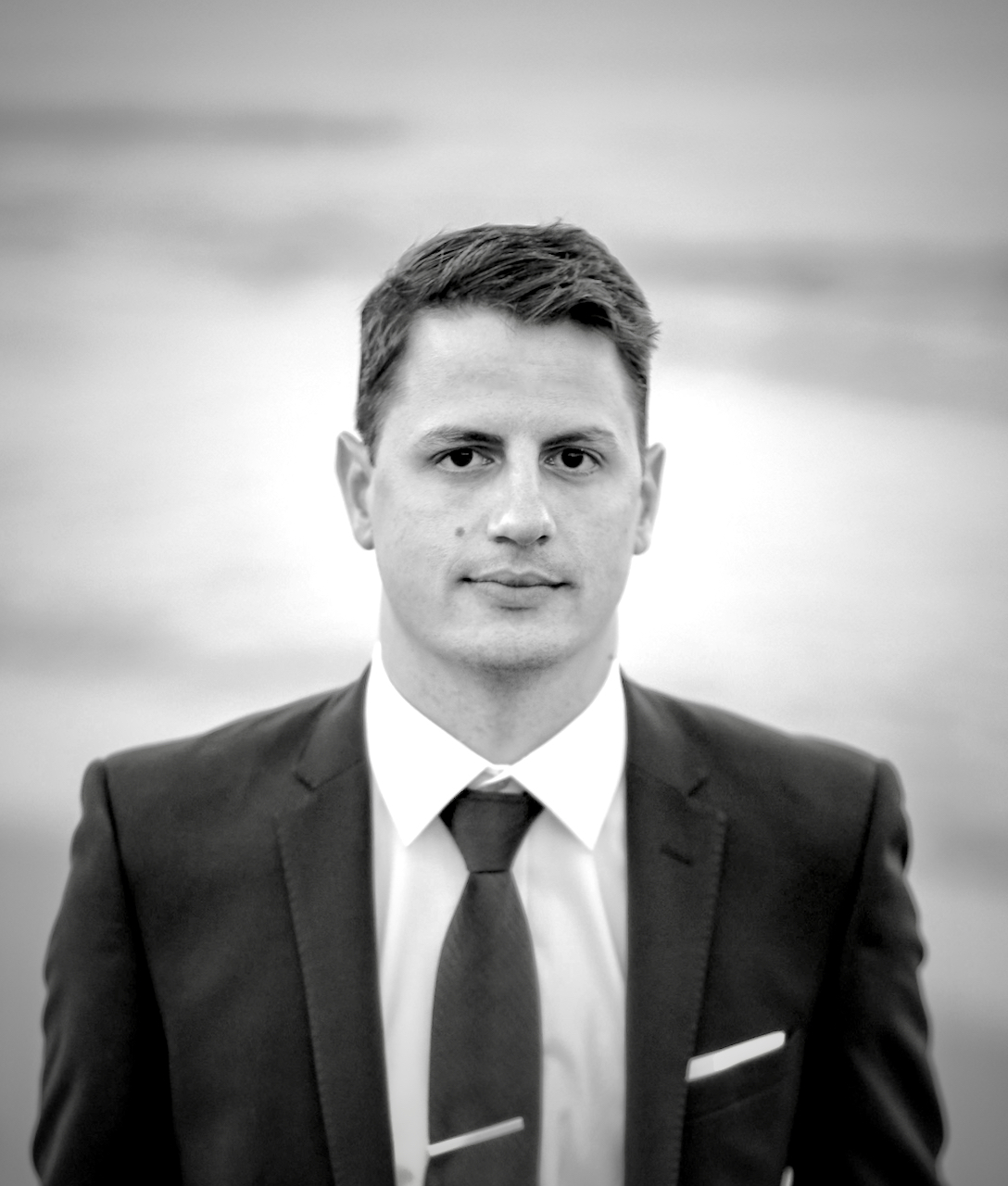}}]{Nick Cvetojevic} received his PhD from Macquarie University, Sydney, Australia, in 2013. From 2013 to 2016, he worked as a post-doctoral researcher on astrophotonic technologies jointly appointed between The University of Sydney and the Australian Astronomical Observatory, where he was involved with the development of novel integrated photonic devices for astronomical interferometry. He led the commissioning of a photonic nulling interferometer module of the SCExAO instrument at the Subaru Telescope, as well as collaborating on on-sky demonstrations of photonic-spectroscopy and efficient telescope injection into single-mode fiber. Between 2016 and 2019 he worked at Observatoire de Paris, France, leading the development of a fiber optic pupil-remapping interferometer (FIRST), focusing on active on-chip electro-optic control for high-contrast imaging and exoplanet characterization. He currently works at Observatoire de la C\^{o}te d'Azur, in Nice, France, focusing on kernel-phase high-contrast detection techniques and wavefront control, as well as developing Kernel-Nulling photonic components for the next generation of VLTI instruments.    
\end{IEEEbiography}%replace with your info please


\vfill


\end{document}